\documentclass[twocolumn]{aastex63}
\usepackage{CJK}
\usepackage{multirow}
\usepackage{booktabs}
\usepackage{rotating}

\newcommand{\wise}{{\textit{WISE~}}}

\newcommand{\wname}{W2246$-$0526}

\shorttitle{Radio activity in W2246$-$0526}
\shortauthors{Fan et al.}

\begin{document}
\begin{CJK*}{UTF8}{gbsn}

\title{The hyperluminous, dust-obscured quasar W2246$-$0526 at $z=4.6$: detection of parsec-scale radio activity}

\correspondingauthor{Lulu Fan, Jun Yang}
\email{llfan@ustc.edu.cn, jun.yang@chalmers.se}

\author[0000-0003-4200-4432]{Lulu Fan (范璐璐)}

\affiliation{CAS Key Laboratory for Research in Galaxies and Cosmology, Department of Astronomy, University of Science and Technology of China, Hefei 230026, China}

\affil{School of Astronomy and Space Sciences, University of Science and Technology of China, Hefei, Anhui 230026, People's Republic of China}

\affil{Shandong Provincial Key Lab of Optical Astronomy and Solar-Terrestrial Environment, Institute of Space Science, Shandong University, Weihai, 264209, People's Republic of China}

\author[0000-0002-5519-0628]{Wen Chen (陈文)} 
\affiliation{Yunnan Observatories, Chinese Academy of Sciences, 396 Yangfangwang, Guandu District, Kunming, 650216, P. R. China}
\affil{University of Chinese Academy of Sciences, No.19(A) Yuquan Road, Shijingshan District, Beijing, 100049, P.R.China}

\author[0000-0003-4341-0029]{Tao An (安涛)} 
\affiliation{Key Laboratory of Radio Astronomy, Shanghai Astronomical Observatory, 80 Nandan Road, 200030, Shanghai, China}

\author[0000-0001-9969-2091]{Fu-Guo Xie (谢富国)} 
\affiliation{Key Laboratory for Research in Galaxies and Cosmology, Shanghai Astronomical Observatory, Chinese Academy of Sciences, 80 Nandan Road, Shanghai 200030, China}

\author[0000-0002-2547-0434]{Yunkun Han (韩云坤)}
\affiliation{Yunnan Observatories, Chinese Academy of Sciences, 396 Yangfangwang, Guandu District, Kunming, 650216, P. R. China}

\author[0000-0002-7821-8873]{Kirsten K. Knudsen}
\affiliation{Department of Space, Earth and Environment, Chalmers University of Technology, Onsala Space Observatory, SE-439 92 Onsala, Sweden}

\author[0000-0002-2322-5232]{Jun Yang (杨军)}
\affiliation{Department of Space, Earth and Environment, Chalmers University of Technology, Onsala Space Observatory, SE-439 92 Onsala, Sweden}

\begin{abstract}

WISE J224607.56$-$052634.9 (\wname{}) is a hyperluminous  ($L_{\rm bol}\approx 1.7\times 10^{14}~L_\odot$), dust-obscured and radio-quiet quasar at redshift $z=4.6$. It plays a key role in probing the transition stage between dusty starbursts and unobscured quasars in the co-evolution of galaxies and supermassive black holes (SMBHs). To search for the evidence of the jet activity launched by the SMBH in \wname{}, we performed very long baseline interferometry (VLBI) observations of its radio counterpart with the European VLBI Network (EVN) plus the enhanced Multi Element Remotely Linked Interferometer Network (e-MERLIN) at 1.66~GHz and the Very Long Baseline Array (VLBA) at 1.44 and 1.66~GHz. The deep EVN plus e-MERLIN observations detect a compact (size $\leq32$~pc) sub-mJy component contributing about ten percent of its total flux density, which spatially coincides with the peak of dust continuum and [C~II] emissions. Together with its relatively high brightness temperature ($\geq8\times10^{6}$~K), we interpret the component as a consequence of non-thermal radio activity powered by the central SMBH, which likely originates from a stationary jet base. The resolved-out radio emission possibly come from a diffuse jet, quasar-driven winds, or both, while the contribution by star formation activity is negligible. Moreover, we propose an updated geometry structure of its multi-wavelength active nucleus and shed light on the radio quasar selection bias towards the blazars at $z>4$.

\end{abstract}


\section{Introduction}

Under the framework of galaxy evolution through major mergers \citep[e.g.][]{Hopkins2008,Somerville2015},  dust-obscured quasars, hosting both intense star formation and active galactic nucleus (AGN) activities, represent the key transition stage between dusty starbursts and unobscured quasars. Therefore, they are the excellent candidates for studying the co-evolution of the massive galaxies and their central SMBHs. By using NASA's {\it Wide-field Infrared Survey Explorer} (\wise) mission \citep{Wright2010}, a new population of hyper-luminous, dust-obscured galaxies (so-called Hot DOGs) have been discovered \citep{Eisenhardt2012, Wu2012}, which have then been revealed as the extremely luminous ($L_{\rm bol}\approx 10^{13.0-14.5}~L_\odot$), merger-driven, heavily obscured quasars at high redshifts ($z\sim1-4.6$) by the follow-up multi-wavelength studies \citep[e.g.,][]{Tsai2015,Piconcelli2015,Assef2016,Ricci2017,Fan2016a,Fan2016b,Fan2017,Fan2018a,Zappacosta2018}.

\wname{}, a Hot DOG at $z\sim4.6$, is extremely luminous and massive, with a bolometric luminosity $L_{\rm bol}\approx 1.7\times 10^{14}L_\odot$ and a stellar mass $M_\star=4.3\times10^{11}{\rm M_\odot}$ \citep{Fan2018b}. It hosts a central SMBH with the black hole mass $M_{\rm BH}\sim 4\times10^9 M_\odot$, accreting at a super-Eddington ratio, $\lambda_{\rm Edd}=2.8$ \citep{Tsai2018}.  According to the {\it HST} near-IR image, \wname{} is likely merging with its close companion galaxy \citep{Fan2016a}. With ALMA [C II] spectra and dust continuum imaging, \wname{} is confirmed to be accreting from at least three companion galaxies \citep{Diaz-santos2018}. Spatially resolved ALMA [C II] observations have shown that it is unstable in terms of the energy and momentum that are being injected into the ISM, strongly suggesting that the gas is being blown away isotropically \citep{Diaz-santos2016}.  All these results are consistent with the merger-driven evolutionary scenario, suggesting that \wname{} is in a key transitionary stage of massive galaxy evolution and on the road to becoming an unobscured quasar. 

Free of obscuration, radio wave has the advantage that it can penetrate deep into the centers of these dusty host galaxies. Recently, \citet{Patil2020} presented the sub-arcsecond-resolution imaging of a mid-infrared and radio-selected sample with 155 heavily obscured, ultra-luminous quasars \citep{Lonsdale2015}, which revealed the compact radio structure in the majority of the sources in their sample. Utilizing the EVN, \citet{Frey2016} performed the VLBI observations of the radio counterparts of four Hot DOGs at $z\sim2$. All the four Hot DOGs were successfully detected at 1.7 GHz in the EVN images. The VLBI imaging results support that radio jets have already  started in the transition phase where starburst and AGN activities coexist. The previous work made by \citet{Frey2016} suggests that Hot DOGs with radio jets represent the earliest stage of the radio AGN evolution \citep[e.g.,][]{An2012,Patil2020}.

\wname{} has a faint radio counterpart with a flux density of $0.62\pm0.15$~mJy at 1.33~GHz  (see the left panel of Fig. \ref{fig:evn}) in the Very Large Array (VLA) Faint Images of the Radio Sky (FIRST) Survey \citep{Becker1995}. We estimate that star formation activity can only contribute a small fraction of the total radio flux density, while AGN-related activity should dominate the radio emission. Here we report our new results of the high-resolution EVN plus e-MERLIN and the VLBA observations of \wname{}. Throughout this work we assume a standard, flat ${\rm \Lambda}$CDM cosmology \citep[see][]{komatsu2011}, with $H_0 = 70$ km~s$^{-1}$~Mpc$^{-1}$, $\Omega_M = 0.3$, and $\Omega_\Lambda = 0.7$. The spatial scale at the distance of \wname{} is $\sim$6.5 pc per milliarcsecond (mas).

\section{Observations and data reduction} \label{sec:data}

\begin{table*}[htbp]
\centering
\caption{Summary of the VLBI observations. The two-letter antenna codes are explained in Appendix \ref{sec:antcode}. }
\label{tab1}
\small
\centering
\begin{tabular}{cccc}
\hline
Date     & $\nu_{obs}$ & Time   & Participating radio telescopes                                    \\
(yymmdd) & (GHz)       &(hours)  &                                                                   \\
\hline
190525   & 1.66        & 7.0 & JB, EF, WB, MC, O8, UR, TR, HH, SV, ZC, IR, SR, CM, DA, KN, PI, DE  \\
190819   & 1.44, 1.66        & 8.0 & BR, FD, HN, KP, LA, MK, NL, OV, PT, SC   \\
\hline
\end{tabular}
\end{table*}

We observed \wname{} at L band with the EVN plus e-MERLIN on 2019 May 25 and the VLBA on 2019 Aug 19. The observation configurations are summarized in Table~\ref{tab1}. 

The EVN plus e-MERLIN observations (project code: EY034) were performed at a central frequency of 1.66~GHz. The data recording rates were 1024 Mbps (16 sub-bands, dual polarization, 16 MHz per sub-band, 2-bit quantization) at the EVN stations and 512 Mbps (2 sub-bands, dual polarization, 64 MHz per sub-band, 2-bit quantization) at the e-MERLIN stations (CM, DA, KN, PI, DE). The data correlation was done by the EVN software correlator \citep[SFXC,][]{Keimpema2015} at JIVE (Joint Institute for VLBI, ERIC) using standard correlation parameters of continuum experiments.

The VLBA observations (project code: BF130) were carried out at both 1.44 and 1.66 GHz simultaneously with its broad-band L-band receivers. The observations had a total recording rate of 2048 Mbps: 4 sub-bands, 128 MHz per sub-band, dual polarization, 2 bit quantization, 1024 Mbps at 1.44 GHz and 1024 Mbps at 1.66 GHz. The data were correlated by the distributed FX software correlator \citep[DiFX,][]{Deller2007} at the NRAO (National Radio Astronomy Observatory) using the normal correlation parameters of continuum experiments (1s integration time and 0.5 MHz frequency resolution).    

All these observations were carried out in the phase-referencing mode. The compact radio source J2248$-$0541 \citep{Beasley2002} had a correlation amplitude of $\sim$0.13~Jy at 2.3 GHz on the long baselines of the VLBA on 2017 May 31 \citep{Charlot2020}, and was used as the phase referencing calibrator. In both EVN plus e-MERLIN and VLBA experiments, we used a nodding cycle time of $\sim$4 minutes. The coordinate of J2248$-$0541 is RA~(J2000) $= 22^{\rm h}48^{\rm m}00\fs0806$ and Dec.~(J2000) $= -05\degr41\arcmin18\farcs219$ in the source catalog of GSFC 2015a obtained from the Goddard Space Flight Center VLBI group \footnote{\url{http://astrogeo.org/vlbi/solutions/rfc_2015a/}}. Compared to its optical position (with an uncertainty of 0.5~mas) provided by the second data release of the \textit{Gaia} \citep{Brown2018}, the VLBI position has an offset: $-$0.9 mas in RA and $-$1.6 mas in Dec. The angular distance between the target and calibrator is about 0.53$\degr$. 

The visibility data were calibrated using the NRAO Astronomical Image Processing System \citep[\textsc{aips} version 31DEC19,][]{Greisen2003} software package. \textsc{A priori} amplitude calibration was performed with the system temperatures and the antenna gain curves provided by each telescope. In the case of missing these calibration data, the nominal system equivalent flux densities recorded in the EVN status table \footnote{\url{http://old.evlbi.org/evlbi/e-vlbi_status.html}} were used. The ionospheric dispersive delays were corrected according to the maps of total electron content obtained from the Global Positioning System (GPS) satellite observations. Phase errors due to the antenna parallactic angle variations were calibrated. After instrumental phase and delay correction were carried out via fringe-fitting on a short scan of the fringe finder 3C~454.3, the global fringe-fitting and bandpass calibration were performed. In the case of the VLBA data, we reduced the two 128 MHz sub-bands per polarization separately because of their large separation in frequency ($\sim$220~MHz). To get the self-calibration solutions at the high-frequency solution, we split each 128~MHz sub-band into eight 16~MHz sub-bands at each frequency.    

The calibrator J2248$-$0541 was imaged using iterations of the \textsc{clean}  and self-calibration (Stokes \textit{I}) in the software package \textsc{difmap} \citep[version 2.5e, ][]{Shepherd1994}, fringe-fitting and self-calibration (Stokes \textit{RR} and \textit{LL}) with the input of the new \textsc{clean} image in \textsc{aips}. In each iteration, both the phase and amplitude self-calibration solutions were also applied to the target source, and the phase errors due to the calibrator structure were removed. The calibrator had a one-sided core--jet structure with total flux densities of 0.40$\pm$0.02~Jy at 1.44 GHz and 0.33$\pm$0.02 Jy at 1.66 GHz. We used the jet base, the most compact and brightest component, as the reference point in the phase-referencing calibration. After three iterations, the deconvolved Stokes $I$ map using natural weighting reached an image noise level of $\sim$90~$\mu$Jy~beam$^{-1}$ in these VLBI images. Thanks to the quite small angular separation (0.53$\degr$) between the calibrator and the target, the phase-referencing calibration worked at a reasonable accurate level. In the residual map of the target source using natural weighting, there are no $>$5$\sigma$ systematic errors (noise peaks, stripes, and rings). We plot the EVN and VLBA images of the phase-referencing calibrator J2248$-$0541 in Appendix \ref{sec:cal_img}.

\section{VLBI imaging results}\label{sec:results}

\begin{figure*}
\centering
\plotone{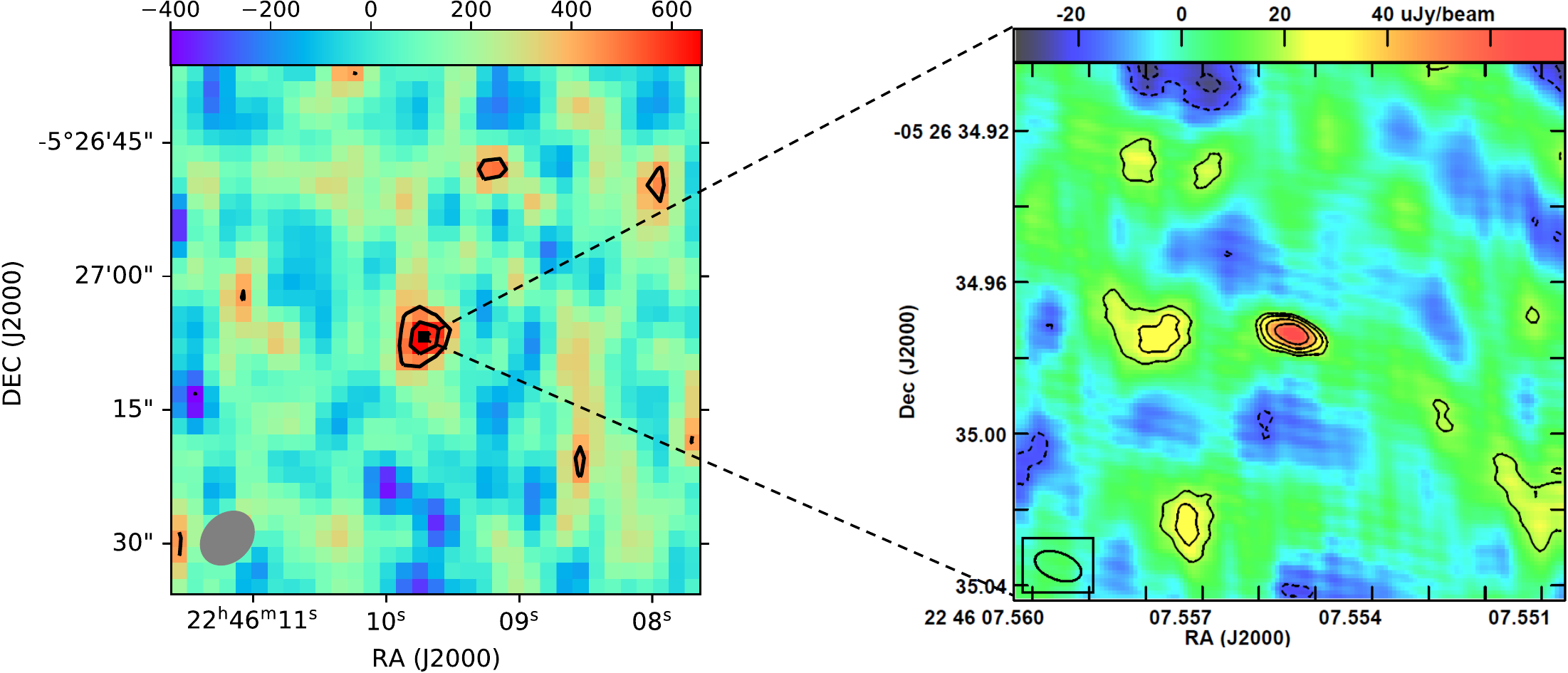}
\caption{
Radio observations of \wname{}. {\it Left:} The $1'\times 1'$ FIRST cutout image of \wname{}. The contours are at 2.5 and 3.5 $\sigma$ level. The beam size is shown in the bottom left-hand corner. {\it Right:} The EVN plus e-MERLIN image of the compact component in \wname{} at 1.66~GHz. The image size is $2''\times2''$. The contours start at 18.4~$\mu$Jy~beam$^{-1}$ ($2.5\sigma$) and increase by factors of $-$1.4, $-$1, 1, 1.4, 2 and 2.8. Only the central peak feature has a brightness (66~$\mu$Jy~beam$^{-1}$) of $>5\sigma$ and thus can be firmly identified as an astrophysical source. The naturally-weighted synthesized beam has dimensions of 12.7~mas $\times$ 6.3~mas (82.6~pc $\times$ 41.0~pc) at a position angle of 72 degrees and is plotted in the bottom left-hand corner.
}
\label{fig:evn}
\end{figure*}

The image of the EVN plus e-MERLIN observations of \wname{} is shown in the right panel of Fig.~\ref{fig:evn}. We detect a single unresolved source co-located with the position of \wname{} without self-calibration. The component has a peak brightness of 66~$\mu$Jy~beam$^{-1}$. Compared to the image noise level 7.4~$\mu$Jy~beam$^{-1}$ in the residual map, the detection has a signal to noise ratio (SNR) of 8.9. We also fit the visibility data in the ($u$, $v$) plane with a circular Gaussian model in \textsc{difmap} and measured the formal uncertainties via setting the reduced $\chi^2=1$. The component has a total flux density of 75$\pm$9~$\mu$Jy and an angular size (FWHM: the full width at half-maximum) of 0.3$\pm$4.9~mas, indicating an unresolved source structure of $\leq$4.9~mas. Using the radio core of the phase-referencing calibrator as the reference point, we carried out differential astrometry and then derived a position for \wname{}, RA~$=22^{\rm h}46^{\rm m}07\fs55542$ ($\sigma_{\rm ra}=0.8$~mas) and  Dec$=-05\degr26\arcmin34\farcs9739$ ($\sigma_{\rm dec}=0.6$~mas).  Because of the detection has a relatively low SNR, these formal uncertainties dominate the associated error budgets. The empirical 1$\sigma$ systematic errors are small, five percent for the total flux density and $\sim$0.3 mas for the position due to the residual phase errors of the phase-referencing calibration. We also searched for the more extended emission via trying a significant Gaussian taper of the visibility data with a value of 0.5 at 2 Mega-wavelength or excluding the long-baseline data, however we did not detect any diffuse component in a field of $2\times2$~arcsec$^{2}$. 

The core brightness temperature can be estimated as \citep[e.g.,][]{Condon1982},
\begin{equation}
T_{\rm b} = 1.22\times10^{9}\frac{S_\mathrm{obs}}{\nu_\mathrm{obs}^2\theta_\mathrm{size}^2}(1+z),
\label{eq1}
\end{equation}
where $S_\mathrm{obs}$ is the observed total flux density in mJy, $\nu_\mathrm{obs}$ is the observing frequency in GHz, $\theta_\mathrm{size}$ is the FWHM of the circular Gaussian model in mas, and $z$ is the redshift. Because the component is unresolved, we take its $1\sigma$ error as an upper limit of its angular size. Thus, the component has $T_{\rm b}\geq8\times10^{6}$ K.  

With the input of the accurate position from the 1.66-GHz EVN plus e-MERLIN detection, we could also gain a marginal detection of \wname{} in the VLBA observations, SNR=3.2 at~1.44 GHz. This is expected since the VLBA observations have a relatively higher image noise, 17~$\mu$Jy~beam$^{-1}$ at 1.44~GHz. The total flux density is $66\pm18$~$\mu$Jy at 1.44~GHz. At 1.66~GHz, we failed to achieve a $\geq3\sigma$ detection of the component because of the limited image sensitivity 23~$\mu$Jy~beam$^{-1}$. We plot the VLBA images of \wname{} at 1.44~GHz and 1.66~GHz in Appendix \ref{sec:vlba_img}.

\section{Discussion} \label{sec:discussion}

\subsection{The origin of radio emission in W2246-0526}\label{subsec:origin}

The EVN plus e-MERLIN observations of \wname{} at 1.66~GHz detects an unresolved faint component with a size of $\leq$32~pc and a radio luminosity of $L_{\rm R}=\nu L_{\nu}(1+z)^{-1-\alpha}= 4.7\times10^{40}$~erg~s$^{-1}$, assuming the spectral index $\alpha=0$. Its position coincides with the peaks of the ALMA 212~$\mu$m dust continuum, [C~II] emission and the {\it HST} near-IR images \citep{Diaz-santos2016,Diaz-santos2018,Fan2016a}. The observed high brightness temperature ($\geq8\times10^{6}$ K) cannot result from  thermal emission, which generally has a $T_{\rm b}<10^6$~K.

The VLBI component cannot be explained as a young radio supernova in the nucleus because its radio luminosity is two orders of magnitude higher than the known maximum luminosity ($L_{\rm R}\sim5\times10^{38}$~erg~s$^{-1}$) of young radio supernovae \citep[e.g.][]{Weiler2002}. Instead, it may be naturally interpreted as the radio activity powered by the central accreting SMBH: a stationary jet base, a young out-moving jet, or a pc-scale radio-emitting outflow. However, due to the limited image resolution and the absence of a spectral index measurement, we can't distinguish among these possibilities. If it has a flat radio spectrum, it would be interpreted as the partially self-absorbed jet base. On the other hand, if it has a steep spectrum, it might result from either the out-moving jet or the pc-scale outflow. However, considering the high value of the intrinsic emission frequency in the rest frame ($\nu_{\rm int} = \nu_{\rm obs} (1+z) = 9.3$~GHz), the flat-spectrum radio emission more likely have the compact structure and the high brightness temperature at the high frequency.

In order to constrain the spectral index on scales of a few arcseconds, we check the VLA Sky Survey \citep[VLASS;][]{Lacy2020} cutout image\footnote{http://cutouts.cirada.ca} of \wname{}, which provides a non-detection and a $3\sigma$ upper limit at 3~GHz, $S_{3{\rm GHz}} < 0.4$~mJy. Combining with the FIRST flux density, we can constrain the spectral index on scales of a few arcseconds ($\alpha < -0.7$).  Thus, we assume a steep radio spectrum with a spectral index $\alpha = -0.8$ to derive the flux density at 1.4~GHz from EVN plus e-MERLIN flux density at 1.66 GHz, which is only  86~$\mu$Jy, 14\% of the FIRST flux density observed on 2011 Mar 19. The intrinsic time interval between FIRST and our observation is $8 (1+z)^{-1} \approx 1.5$~yr. Typically, the radio luminosity variation is weak among non-blazar sources on the short timescale. Thus, the flux difference is most likely due to some low-brightness diffuse radio emission which is completely resolved out in the EVN image. The extended radio emission can come from intense star formation or AGN activities (jet or outflow). The derived star formation rate (SFR) of \wname{} is about 500 M$_\odot$~yr$^{-1}$ based on the IR luminosity \citep{Fan2018b,Diaz-santos2018}. The corresponding radio power can be estimated from the conversion relation \citep{Murphy2011}:  
\begin{equation}
\left(
\frac{{\rm P}_{1.4 {\rm GHz}}}{{\rm erg~s^{-1}~Hz^{-1}}}
\right) = 1.6\times10^{28} 
\left(
\frac{{\rm SFR}_{1.4 {\rm GHz}}}{{\rm M}_\odot~{\rm yr}^{-1}}
\right)
\label{eq2}
\end{equation}
The corresponding flux density S$_{\rm 1.4GHz}$ can be calculated by the equation 
\begin{equation}
S_{\rm 1.4~GHz} = \frac{{\rm P}_{1.4{\rm GHz}}}{4\pi{ D_{\rm L}^2}(1+z)^{-1-\alpha}}
\label{eq3}
\end{equation}
where $D_{\rm L}$ is the luminosity distance of \wname{}.
We get $S_{\rm 1.4GHz}$ = 5.3$\,\mu$Jy, which is only 1\% of the extended radio emission. Therefore the majority of the radio emission resolved out with EVN plus e-MERLIN could be associated with the AGN-related activity (jet and/or outflow). 

The significantly over-resolved VLBI radio structure in \wname{} is frequently seen among these IR to sub-mm luminous but radio-quiet (or weak) sources \citep[e.g.][]{Frey2016, Chen2020}. Compared to the Hot DOGs at $z \sim 2-3$ observed by \citet{Frey2016}, \wname{} represents the more energetic case locating at the higher redshift. The VLBI-detected component is more compact and likely includes a compact radio core. The best nearby ($z < 0.3$) analogy to \wname{} might be the optically luminous quasar PDS~456 at $z=0.184$. PDS~456 has a $\sim10^9~ M_\odot$ black hole accreting at the Eddington rate, both of which are similar to those of our target. The VLBI observations of PDS~456 revealed a quite complex radio nucleus consisting of some very extended emission and a faint jet, including a candidate radio core at 1.66~GHz \citep{Yang2020}. At the higher frequency $>$5~GHz, only the candidate radio core would be detectable because of its partially optically thick radio spectrum \citep{Yang2019, Yang2020}.

\subsection{The heavily dust-obscured, radio-quiet quasar at high redshift}

\begin{figure}
\plotone{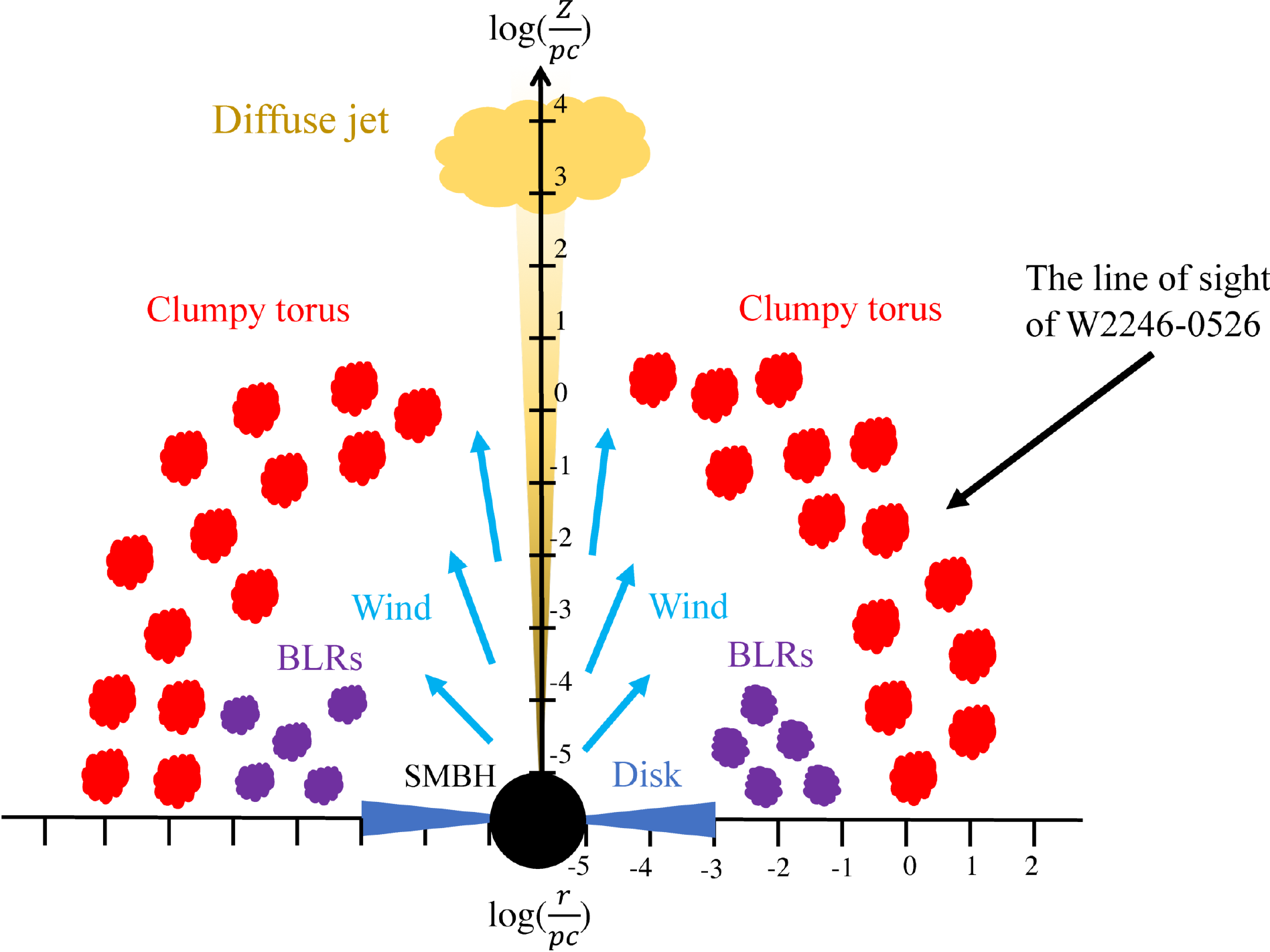}
\caption{A cartoon of the proposed structure in the inner region of \wname{}, based on the AGN physical model in \citet{Ramos2017}. Radio emissions in \wname{} mainly come from a stationary jet base or a newly-emerging moving-out jet (the unresolved compact core component), and the quasar-driven radio-emitting winds and/or a diffuse jet (the extended radio emission). The line of sight of \wname{} and the possible jet are misaligned. \wname{} is optically faint due to the heavy dust obscuration (the covering factor of dust torus is close to 1), and thus can be missed by optical surveys, such as the Sloan Digital Sky Survey (SDSS). \label{fig:cartoon}}
\end{figure}

According to the spectral energy distribution (SED) modeling on the rest-frame UV/optical-to-far-IR of \wname{} \citep{Fan2018b}, we estimated the unattenuated flux density at 4400 \AA~ (B band) $f_{\rm B} = 29$~$\mu$Jy. The radio flux density at 5 GHz $f_{\rm R}$ is estimated by extrapolating from  $f_{\rm 1.4 GHz}$. The radio loudness $R = f_{\rm R}/f_{\rm B}$ of \wname{} is 1.1 by taking the EVN value $f_{\rm 1.4 GHz} = 86$~$\mu$Jy, or 7.7 by taking the FIRST value $f_{\rm 1.4 GHz} = 620$~$\mu$Jy. In either case, the radio loudness $R$ is less than 10. Therefore \wname{} will be classified as a radio-quiet quasar \citep{Kellermann1989}. The radio loudness $R$ has been found to be strongly inversely correlated with the accretion ratio $\lambda$ \citep{Ho2002,Sikora2007}. Based on the $R-\lambda$ relation given by \citet{Ho2002}, the expected radio loudness is 1.1 from  the Eddington ratio of \wname{} $\lambda_{\rm Edd} = 2.8$ \citep{Tsai2018}, which is in good agreement with our above estimation. 

\wname{} has been suggested to be blowing out its ISM isotropically, based on the finding of a large uniform velocity dispersion of the [C~II] emission \citep{Diaz-santos2016}. Shocks, which propagate through the host galaxy as a result of quasar-driven winds, can accelerate relativistic electrons in quasar-driven winds and produce the non-thermal synchrotron radio emission with $\nu{L}_{\rm \nu} ~\sim~ 10^{-5}~{L_{\rm AGN}}$ on scales $>~ 0.1$ kpc \citep{Nims2015}. Given the bolometric luminosity of \wname{} (${ L_{\rm bol}}~\sim~1.7\times10^{14}~{L_\odot}$), the expected radio emission will be about $10^{42-43}~{\rm erg~s^{-1}}$, which is generally consistent with the observed radio emission at 1.4 GHz. The origin of the radio emission from quasar-driven winds has been supported by the observed correlation between the radio luminosity and the velocity dispersion of [O III]-emitting ionized gas in low- \citep{Zakamska2014} and high-redshift, radio quiet quasars \citep{Hwang2018}. However, compact, steep-spectrum, low-power jets launched by central SMBHs can also produce the radio emission with the similar luminosity to the observed value of \wname{} \citep{Leipski2006}. Especially, a jet is expected to be formed in \wname{}, which has high gas and dust contents on the scale of 2$\sim$3 kpc \citep{Fan2016b,Diaz-santos2018}, resulting in efficient confinement and formation of diffuse jets. 

As discussed above, quasar-driven winds and/or a diffuse jet can dominate the extended radio emission in \wname{}, while star formation can only contribute a small fraction of $\sim 1\%$. We propose a possible scenario of the inner region of \wname{}, which is shown as a cartoon in Fig. \ref{fig:cartoon}. The heavy obscuration is suggested by the result of infrared SED decomposition, where the covering factor of the clumpy dust torus is close to 1 \citep{Fan2016b}. The heavily obscured dust structure together with a relativistic jet is exactly what had been expected by \citet{Ghisellini2016}, in order to explain the disagreement between the observed and misaligned quasar numbers at $z > 4$ \citep{Volonteri2011}. Radio sources like \wname{} show faint optical emission due to the obscuration, and thus cannot be selected by the match of radio and optical catalogs. Only the line of sight along the jet can observe the unattenuated optical emission from the accretion disc and the broad emission line regions (BLRs). The selection is then biased towards the blazars.   

\section{Summary} \label{sec:summary}

Based on the high-resolution radio observations of the hyperluminous, dust-obscured, quasar \wname{} at $z=4.6$ with the EVN plus e-MERLIN and the VLBA, we achieved a SNR$\sim$9 detection of an unresolved component ($\leq32$~pc) at 1.66~GHz in the nucleus. The pc-scale compact component has a brightness temperature $T_{\rm b}\geq8\times10^{6}$ K, which can be naturally explained as the radio activity powered by the central accreting SMBH: a stationary jet base, a out-moving jet or radio-emitting winds. The flux density of the compact core component (75$\pm$9~$\mu$Jy) only accounts for about ten percent of the known FIRST flux density. Given the SFR$\sim$500 M$_\odot$~yr$^{-1}$ of \wname{},  the corresponding radio flux density from star formation is only 1\% of the extended radio emission that is completely resolved out with EVN plus e-MERLIN. Therefore, the majority of the VLBI-undetected radio emission could be associated with the quasar-driven winds and/or a diffuse jet. The nature of this diffuse radio emission can be tested in the future using current instruments like the VLA (A-config).  Compared to the heavily obscured, hyperluminous quasars at $z \sim 0.4-3$ \citep{Frey2016,Patil2020}, \wname{} represents the most luminous case ($L_{\rm bol} = 1.7\times10^{14}$ erg~s$^{-1}$) locating at the higher redshift, and the VLBI-detected component is more compact.

The radio loudness $R$ is estimated to be 1.1 or 7.7 depending on the choice of the radio flux density of the EVN plus e-MERLIN value or the FIRST value. Thus, \wname{} is a rare, radio-quiet quasar with high-resolution VLBI detection, while most if not all of those VLBI-detected quasars at $z>4.5$ are radio loud \citep{Coppejans2016}. Combining the possible evidence of jet activity and the heavy dust obscuration in \wname{}, we propose a possible scenario of its inner region. The proposed scenario is consistent with the expectation by \citet{Ghisellini2016} in order to explain the disagreement of the observed number among blazars and misaligned radio quasars at $z > 4$.

\begin{acknowledgements}

We thank the anonymous referee for constructive comments and suggestions. LF, TA and FGX acknowledge the support from National Key Research and Development Program of China (Nos. 2017YFA0402703, 2018YFA0404603 and 2016YFA0400704). This work is supported by the National Natural Science Foundation of China (NSFC, grant Nos. 11822303, 11773020, 11421303, 11903079, 11873074 and 11773063), Shandong Provincial Natural Science Foundation, China (JQ201801) and the Strategic Priority Research Program of Chinese Academy of Sciences (Grant No. XDB 41000000). KK acknowledges support from the Knut and Alice Wallenberg Foundation.

The European VLBI Network is a joint facility of independent European, African, Asian, and North American radio astronomy institutes. Scientific results from data presented in this publication are derived from the following EVN project code: EY034. 
e-MERLIN is a National Facility operated by the University of Manchester at Jodrell Bank Observatory on behalf of STFC.
The research leading to these results has received funding from the European Commission Horizon 2020 Research and Innovation Programme under grant agreement No. 730562 (RadioNet).
The National Radio Astronomy Observatory is a facility of the National Science Foundation operated under cooperative agreement by Associated Universities, Inc.
This work made use of the Swinburne University of Technology software correlator, developed as part of the Australian Major National Research Facilities Programme and operated under license.

\facility{EVN, e-MERLIN, VLBA}

\end{acknowledgements}

\appendix

\section{The telescopes participating in the VLBI observations}\label{sec:antcode}

A total of 17 telescopes participated in the deep EVN plus e-MERLIN observations: Jodrell Bank Mk2 (JB), Westerbork (WB, single dish), Effelsberg (EF), Medicina (MC), Onsala (O8), Urumqi (UR), Toru\'n (TR), Hartebeesthoek (HH), Svetloe (SV), Zelenchukskaya (ZC), Irbene (IR), Sardinia (SR), Cambridge (CM), Darnhall (DA), Knockin (KN), Pickmere (PI), and Defford (DE).

The participating telescopes in the VLBA observations were St. Croix (SC), Hancock (HN), North Liberty (NL), Fort Davis (FD), Los Alamos (LA), Pie Town (PT), Kitt Peak (KP), Owens Valley (OV), Brewster (BR) and Mauna Kea (MK).   

\section{EVN and VLBA images of the phase-referencing calibrator}\label{sec:cal_img}

\begin{figure*}
\includegraphics[width=0.32\textwidth]{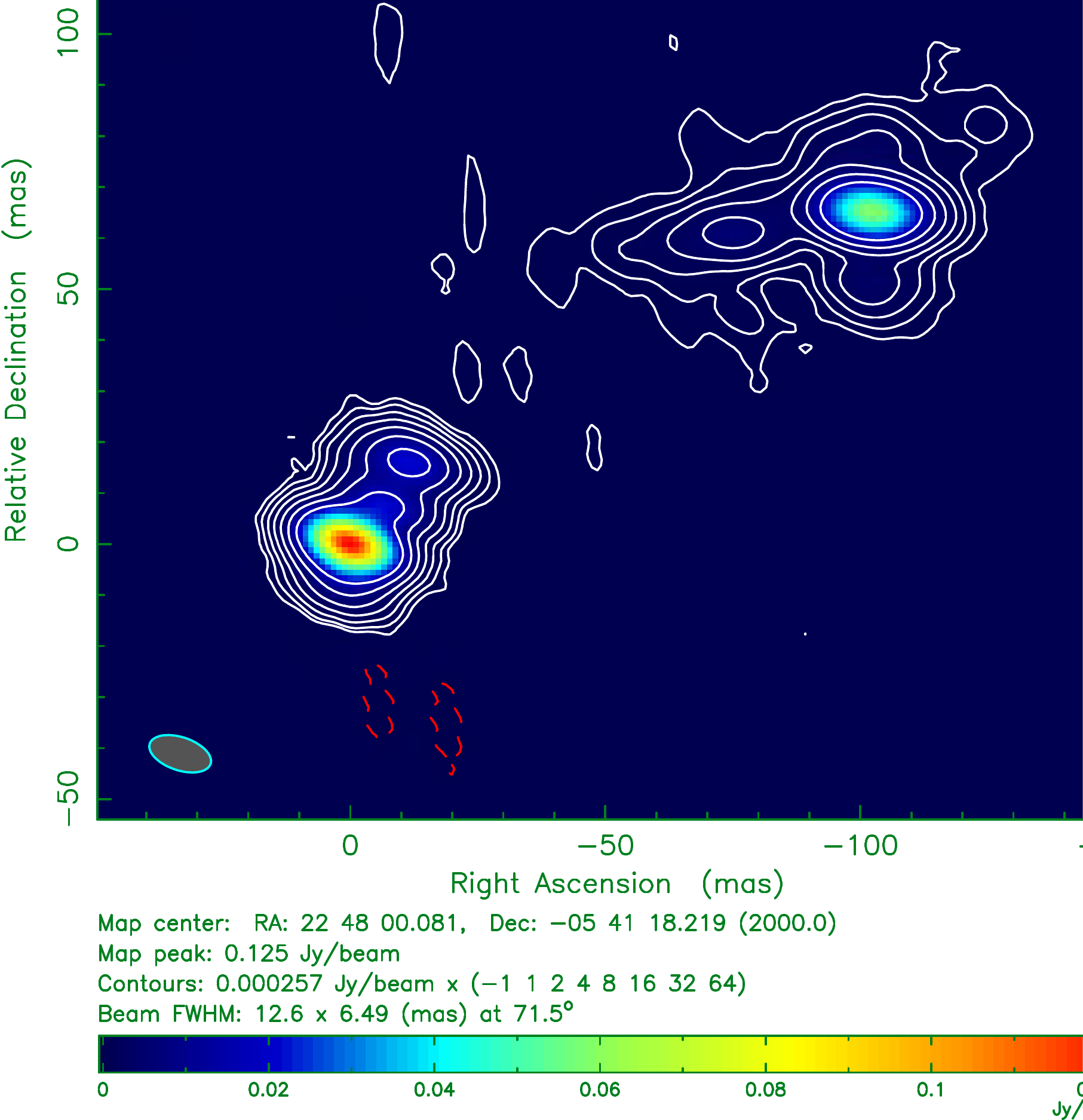}
\includegraphics[width=0.32\textwidth]{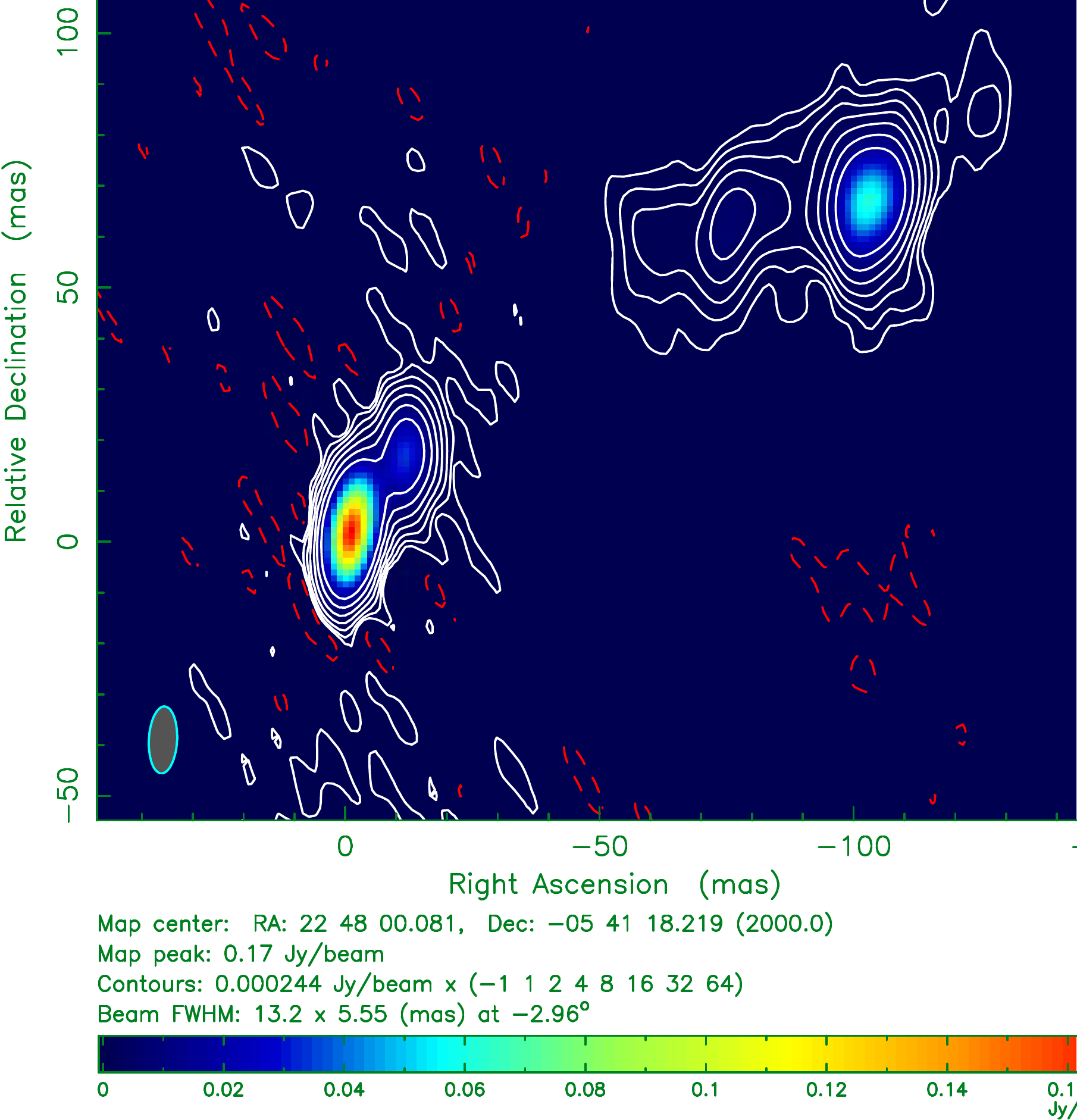}
\includegraphics[width=0.32\textwidth]{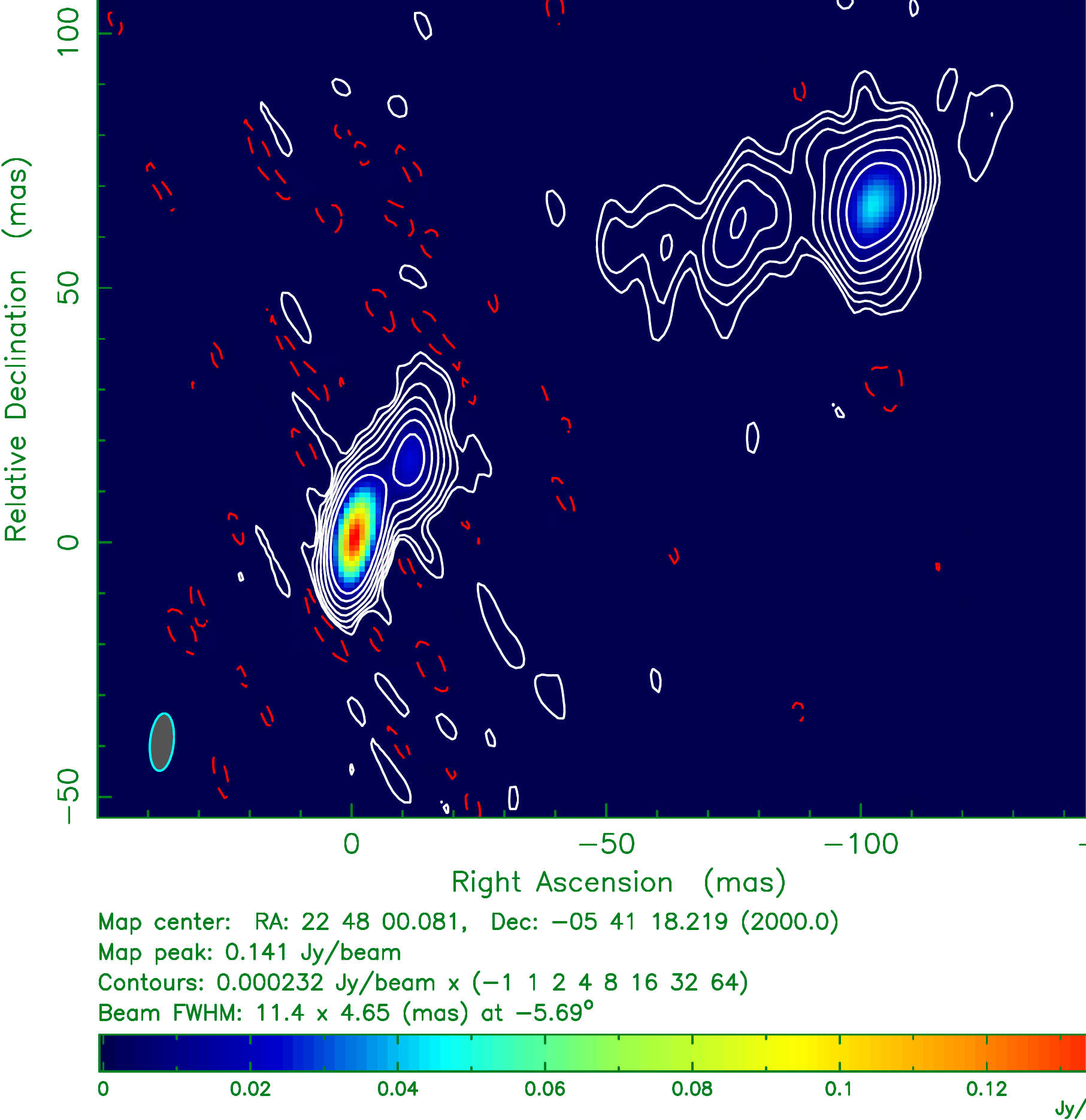}
\caption{The VLBI images of the phase-referencing calibrator J2248$-$0541 at 1.44~GHz by EVN (left panel), 1.44~GHz by VLBA (middle panel) and 1.66~GHz by VLBA (right panel). The first contours are at the 3$\sigma$ level. }
\label{fig:phase_cal}
\end{figure*}

We provide the EVN and VLBA images of the phase-referencing calibrator J2248$-$0541 in Fig. \ref{fig:phase_cal}.

\section{VLBA images of W2246-0526}\label{sec:vlba_img}

\begin{figure*}
\plottwo{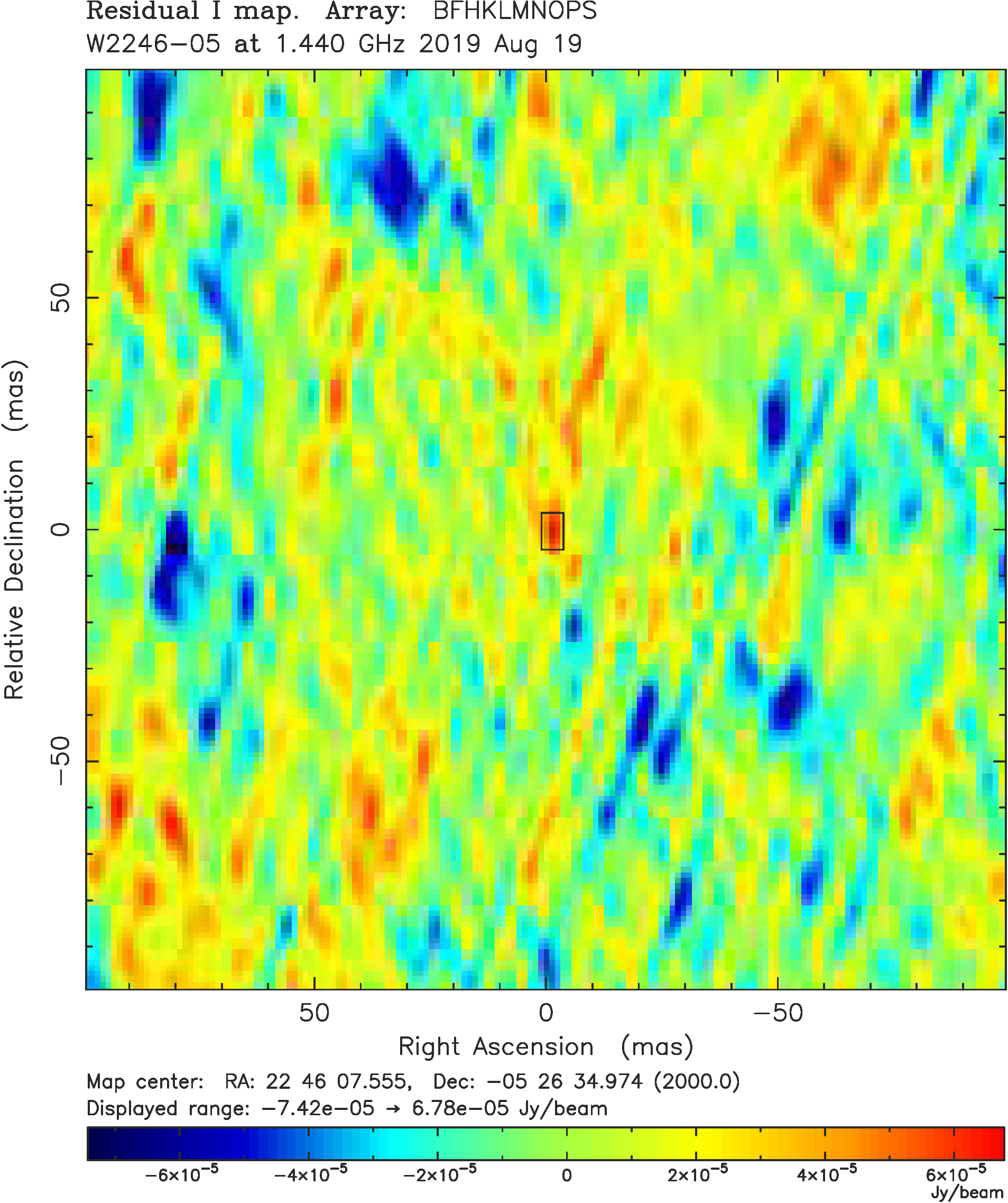}{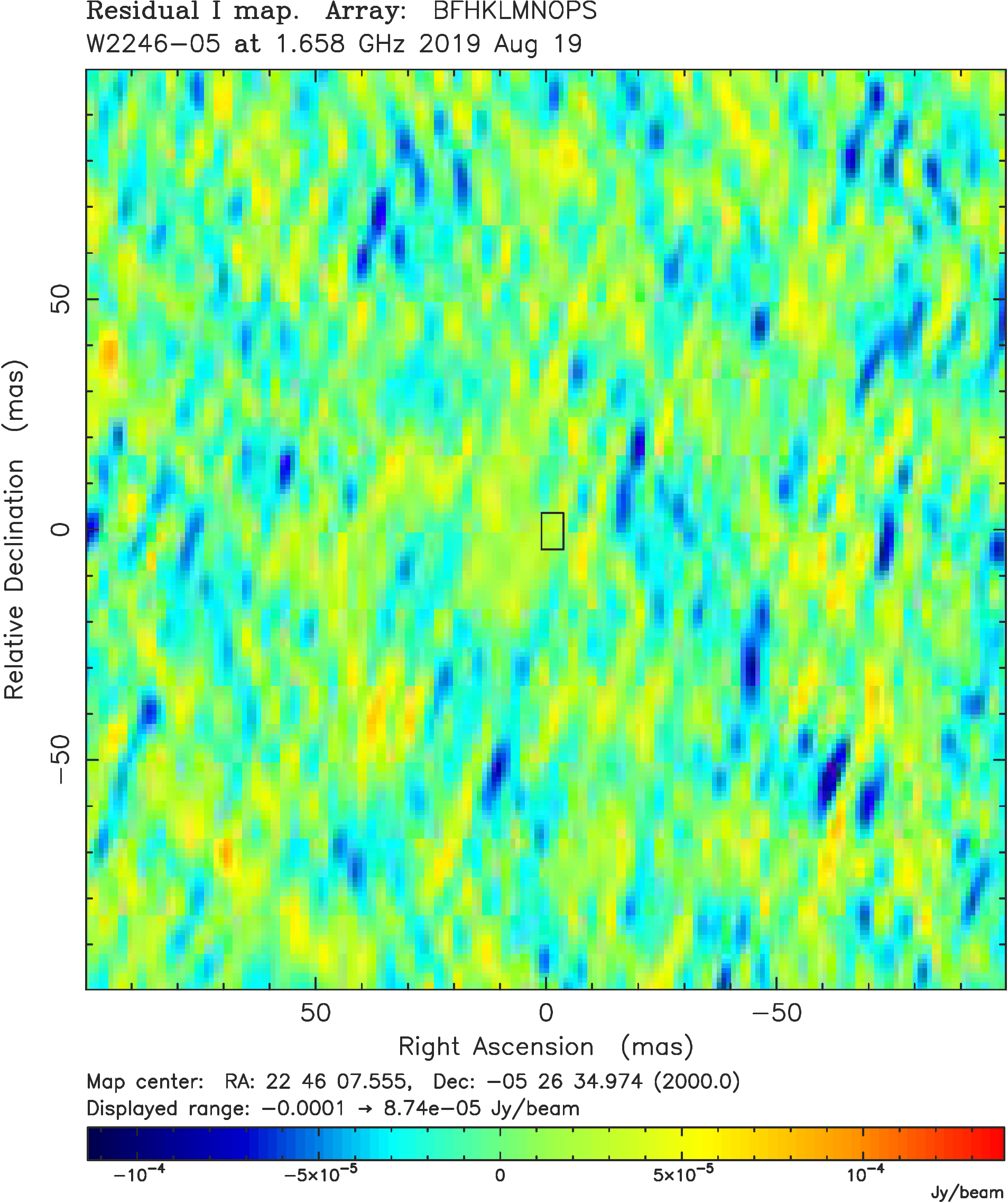}
\caption{The dirty maps of \wname{} observed by the VLBA simultaneously at 1.44~GHz (left panel) and 1.66~GHz (right panel). The rectangles on both panels mark the position of the compact component in \wname{} detected by the EVN plus e-MERLIN. We gain a marginal detection (SNR=3.2) at 1.44~GHz, while fail to achieve a $\geq3\sigma$ detection at 1.66~GHz. }
\label{fig:vlba}
\end{figure*}

We provide the VLBA images of \wname{} in Fig. \ref{fig:vlba}.

\end{CJK*}
\end{document}